\documentclass[aps,prl,amsmath,amssymb,superscriptaddress,showpacs,balancelastpage,10pt,reprint]{revtex4-1}

\usepackage[dvips]{graphicx}
\usepackage{epsfig}
\usepackage[colorlinks=true]{hyperref}
\usepackage{pstricks}

\newcommand{\dotex}{\frac{d}{dt}}
\newcommand{\tr}[1]{\text{Tr}\left(#1\right)}
\newcommand{\trr}[1]{\text{Tr}^2\left(#1\right)}

\newcommand{\RR}{{\mathbb R}}

\newcommand{\ket}[1]{\left| #1 \right\rangle}
\newcommand{\bra}[1]{\left\langle #1 \right|}
\newcommand{\bket}[1]{\left\langle #1 \right\rangle}

\newcommand{\ba}{\text{\bf{a}}}
\newcommand{\Sm}{\boldsymbol{\sigma_{\text{\bf -}}}}
\newcommand{\Sp}{\boldsymbol{\sigma_{\text{\bf +}}}}
\newcommand{\bn}{\text{\bf{n}}}

\newcommand{\GE}{{\vert g \rangle\langle e \vert}}

\newcommand{\Id}{{\mathbb I}}
\newcommand{\Dm}{{\mathcal D}_{m}}

\newcommand{\rhoLR}{{\rho_{\text{\tiny LR}}}}
\newcommand{\psiLR}{\psi_{\text{\tiny LR}}}

\begin{document}

\title{Adaptive low-rank approximation and denoised Monte-Carlo approach\\ for high-dimensional Lindblad equations}

\author{C. Le Bris}
\affiliation{\'Ecole des Ponts and Inria, 6 et 8 avenue Blaise Pascal, 77455 Marne-La-Vall\'ee Cedex 2, France.}\label{I1}
\author{P. Rouchon }
\email{pierre.rouchon@mines-paristech.fr}
\affiliation{Centre Automatique et Syst\`{e}mes, Mines-ParisTech, PSL Research University,  60, bd Saint-Michel, 75006 Paris, France.}\label{I1}
\author{J. Roussel}
\affiliation{\'Ecole des Ponts and Inria, 6 et 8 avenue Blaise Pascal, 77455 Marne-La-Vall\'ee Cedex 2, France.}\label{I1}

\date{\today}


\begin{abstract}
We present a twofold contribution to the numerical simulation of Lindblad equations. First, an adaptive numerical approach to approximate Lindblad equations using low-rank dynamics is described: a deterministic low-rank  approximation of the density operator is computed, and its rank is adjusted dynamically, using an on-the-fly estimator of the error committed when reducing the dimension. On the other hand, when the intrinsic dimension of the Lindblad equation is too high to allow for such  a deterministic  approximation, we combine  classical ensemble averages of
quantum Monte Carlo trajectories and a denoising technique. Specifically, a variance reduction method based upon the consideration of a low-rank dynamics as a control variate is developed. Numerical tests for  quantum collapse and revivals show the efficiency of each approach, along with the complementarity of the two approaches.

\end{abstract}

%
%
%

\maketitle

\section{Introduction}
Lindblad equations are notoriously challenging to simulate numerically. The two categories of approaches are deterministic approaches, on the one hand, and Monte-Carlo approaches \cite{DalibCM1992PRL}, on the other hand. Both categories have their pros and cons. In the former category, the simulation is extremely effective when possible, but a major difficulty lies in the high dimensionality of the ambient space which drastically limits the applicability. In the latter category, dimensionality is not an issue, but the intrinsic noise of stochastic simulations affects the quality of the numerical results. The twofold purpose of this article is to present recent advances in either of the two categories of approaches.

In a previous work~\cite{LeR2013PRA}, the first two authors have presented a possible deterministic approach for the simulation of the Lindblad equation, actually borrowed from similar ideas introduced in~\cite{Handel-mabuchi:JOB2005,Mabuchi:PRA2008} in the context of
quantum filtering.
The approach consists in approximating the evolution of the $n\times n$ density matrix $\rho$  solution to the differential Lindblad equation using a
reduced dynamics on the set of  density matrices of some fixed rank $m \ll
n$. This reduced dynamics  is obtained by  taking  the orthogonal projection  of $\dotex \rho$ onto the tangent space to this set of rank-$m$ matrices  . The clear limitation of the approach lies in the fact that many practical problems are \emph{not} reducible to a low-rank approximation, and further that, even when it is the case, the intrinsic dimensionality of the reduced dynamics is not necessarily known beforehand and may vary in time. So the question of \emph{adjusting} on-the-fly the dimensionality~$m$ of the low rank dynamics immediately arises. As our first contribution in the present article, we describe below an \emph{adaptive} low-rank simulation, the purpose of which is to significantly extend the applicability of the approach introduced earlier in~\cite{LeR2013PRA}. The questions we examine in this work enjoy some similarity with questions arising in computational quantum chemistry, typically for multi-configuration time-dependent Hartree and Hartree-Fock equations~\cite{MR2474331,LubichBIT2014}.

For problems definitely not amenable to deterministic simulation because of their prohibitively high dimensionality (which is indeed the case for many practically relevant problems), stochastic approaches are in order, {\it see}~\cite{DalibCM1992PRL,Moelmer1993,CastiM1995PRL,haroche-raimondBook06}. Although the low-rank dynamics no longer adequately represents the system, a reduced model, simulated deterministically, can however serve as a useful tool for the stochastic simulation of the high dimensional system. We employ the reduced dynamics as  a \emph{control variate} within a variance reduction method applied to the full, high dimensional stochastic system. Our second contribution is to demonstrate the efficiency of such a variance reduction method.

\section{Adaptive low-rank approximation}
We consider throughout this article a Lindblad equation with, for simplicity (and this is by no means a limitation of our methods), a single decoherence operator~$L$,
\begin{equation}\label{eq:lindblad}
    \dotex \rho = -i[H,\rho]  - \tfrac{1}{2} (L^\dag L \rho + \rho L^\dag L) + L \rho L^\dag,
\end{equation}
where $\rho$ is a $n\times n$ non-negative Hermitian matrix with
$\tr{\rho}=1$, $H$ is a $n\times n$ Hermitian matrix and $L$ is a $n\times
n$ matrix. The reduced dynamics derived in~\cite{LeR2013PRA}  approximates, for $n$ large,  the above
dynamics on the set of non-negative Hermitian matrices of rank $m$, $m$
being an integer presumably much
smaller than $n$.
To make the approximation explicit, one introduces a system of  two coupled  differential equations for $U$ and $\sigma$ corresponding to
the generic decomposition~$\rhoLR = U \sigma U^\dag$, where $\sigma$ is a $m\times m$ strictly positive Hermitian matrix,  $U$
a $n\times m$ matrix with $U^\dag U =\Id_m$, and $\Id_m$
denotes the $m\times m$ identity matrix.
That system reads as:
\begin{align}
\dotex U &= - i H U \notag \\
& \quad+(\Id_n-UU^\dag) \left(- \tfrac{1}{2}L^\dag LU +LU\sigma U^\dag L^\dag U\sigma^{-1}\right),
    \label{eq:UH}
    \\
  \dotex \sigma &=  - \tfrac{1}{2} (U^\dag L^\dag L U \sigma + \sigma U^\dag L^\dag L U)
  + U^\dag L U\sigma  U^\dag L^\dag U \notag
  \\&\quad + \tfrac{1}{m}\tr{L^\dag (\Id_n-UU^\dag)L ~U\sigma U^\dag} \Id_m.
    \label{eq:sigH}
\end{align}
Notice that   $H$ only appears in~\eqref{eq:UH} and not in~\eqref{eq:sigH}, a fact that is particularly appropriate when $H$ dominates $L$, in which case~\eqref{eq:sigH} may be understood as a slow evolution as compared to the
 dynamics~\eqref{eq:UH}. Then the projection of the original Lindblad dynamics~\eqref{eq:lindblad} onto the tangent space to
the set $\Dm$ of density matrices of rank $m$  takes the explicit form
 \begin{multline}
\label{eq:LowRankbis}
    \dotex \rhoLR =  -i[H,\rhoLR]  - \tfrac{1}{2} (L^\dag L \rhoLR + \rhoLR L^\dag L) + L \rhoLR L^\dag \\- (\Id_n-P_\rhoLR)L\rhoLR L^\dag (\Id_n-P_\rhoLR) \\ +\tfrac{\tr{L\rhoLR L^\dag(\Id_n-P_\rhoLR)}}{m} P_\rhoLR,
\end{multline}
where the orthogonal projection on the image of $\rhoLR$, $P_\rhoLR=UU^\dag $, only depends on $\rhoLR$.  Notice that the rightmost term of ~\eqref{eq:LowRankbis} allows $\tr{\rhoLR}$ to be preserved in time.
 In the sequel, we denote by $ \mathcal{L}$ the right-hand side of~\eqref{eq:lindblad}, $ \mathcal{L}^\parallel $ that of ~\eqref{eq:LowRankbis}, and by~$\mathcal{L}^\perp=\mathcal{L}-\mathcal{L}^\parallel$.

We have described in details in~\cite{LeR2013PRA} how system~\eqref{eq:UH}-\eqref{eq:sigH} may be efficiently simulated and  then provides an accurate approximation of~\eqref{eq:lindblad} in the case when the rank can be actually reduced. In that work, the rank~$m$ was prescribed \emph{beforehand}. Our purpose here is to explain how the approach can be amended so as to allow for a \emph{dynamical adaptation} of the rank~$m$ of the reduced system.

The adaptation we suggest is based on the evaluation, and update, of the projection error
\begin{multline*}
  	\mathcal{L}^\perp (\rhoLR)= (\mathbb{I}_n-P_\rhoLR)L\rhoLR L^{\dagger} (\mathbb{I}_n-P_\rhoLR) \\
  - \frac{ \tr{L\rhoLR L^{\dagger}(\mathbb{I}_n-P_\rhoLR)} }{m} P_\rhoLR
\end{multline*}
committed when replacing~\eqref{eq:lindblad} by~\eqref{eq:LowRankbis}.

This error may be reduced upon adding one dimension to the $m$-dimensional subspace~${Im} (U)$ associated to the projector~$P_\rhoLR$. The best possible such dimension to add is that for which the projection error is minimal. The rank-$m$ projector~$P_\rhoLR = U U^{\dagger} \in \mathbb{R}^{m \times m}$ is modified into the rank-$m+1$ projector~$P_\rhoLR + Q = U U^{\dagger} + V V^{\dagger} \in \mathbb{R}^{m \times m}$ where~$Q = V V^{\dagger}$, with~$V\in\RR^n$,  is the one-dimensional projector associated to the one dimension added. Denoting by
$$
G = (\mathbb{I}_n-P_\rhoLR) L \rhoLR L^{\dagger} (\mathbb{I}_n-P_\rhoLR),
$$
 a straightforward calculation yields
\begin{multline*}
  	\trr{\mathcal{L}^\perp_{P+Q} (\rhoLR)}
	=
 \| (\mathbb{I}_n-Q) G (\mathbb{I}_n - Q) \|^2
 \\
+ \tfrac{1}{m+1} \trr{(\mathbb{I}_n - Q) G }.
\end{multline*}

One may then prove (see the details in~\cite{RousselMaster2015}) that the directions
\begin{equation}
\label{eq:V}
	V = \hbox{\rm Argmin} _{{V \perp U ,\\ \|V\|=1}} \| \mathcal{L}^\perp_{P_\rhoLR+Q} (\rhoLR) \|
\end{equation}
minimizing the projection error are the eigenvectors of the symmetric matrix~$G$ associated to its largest eigenvalue.
The matrix~$G$ being of large size~$n$, determining its largest eigenvector is challenging computationally. To this end, we notice that
$$
	V^{\dagger} G V
	= \|V^{\dagger} (\mathbb{I}_n-P_\rhoLR) LU \sqrt{\sigma} \|^2,
$$
and that  the range of $(\mathbb{I}_n-P_\rhoLR) LU$ is of dimension less or equal to $m$ and is orthogonal to the range of $\rhoLR$.
Thus, denoting by $\ (\phi_j)_{0\leq j < r}$ an orthonormal basis of the range of~$(\mathbb{I}_n-P_\rhoLR) LU$ with $r \leq m$,  it is sufficient to consider $V$ as linear combination of the $\phi_j$ to  get an eigenvector of~$G$ with largest eigenvalue: $V=\sum_{j=1}^r v_j \phi_j$  where the~$r$-dimensional vector $v$ of component $v_j$ corresponds to the eigenvector with largest eigenvalue of
$$
	K = \Phi^{\dagger} (\mathbb{I}_n-P_\rhoLR) L \rhoLR L^{\dagger} (\mathbb{I}_n-P_\rhoLR) \Phi,
$$
with  $\Phi$ the  $n\times r$ matrix  formed by the $r$  vectors   $\phi_j$. Since~$K$  is of  size~$r\leq m$, this provides an effective  manner to determine the optimum in~\eqref{eq:V}.

\begin{figure}
\centerline{\includegraphics[width=0.8\columnwidth]{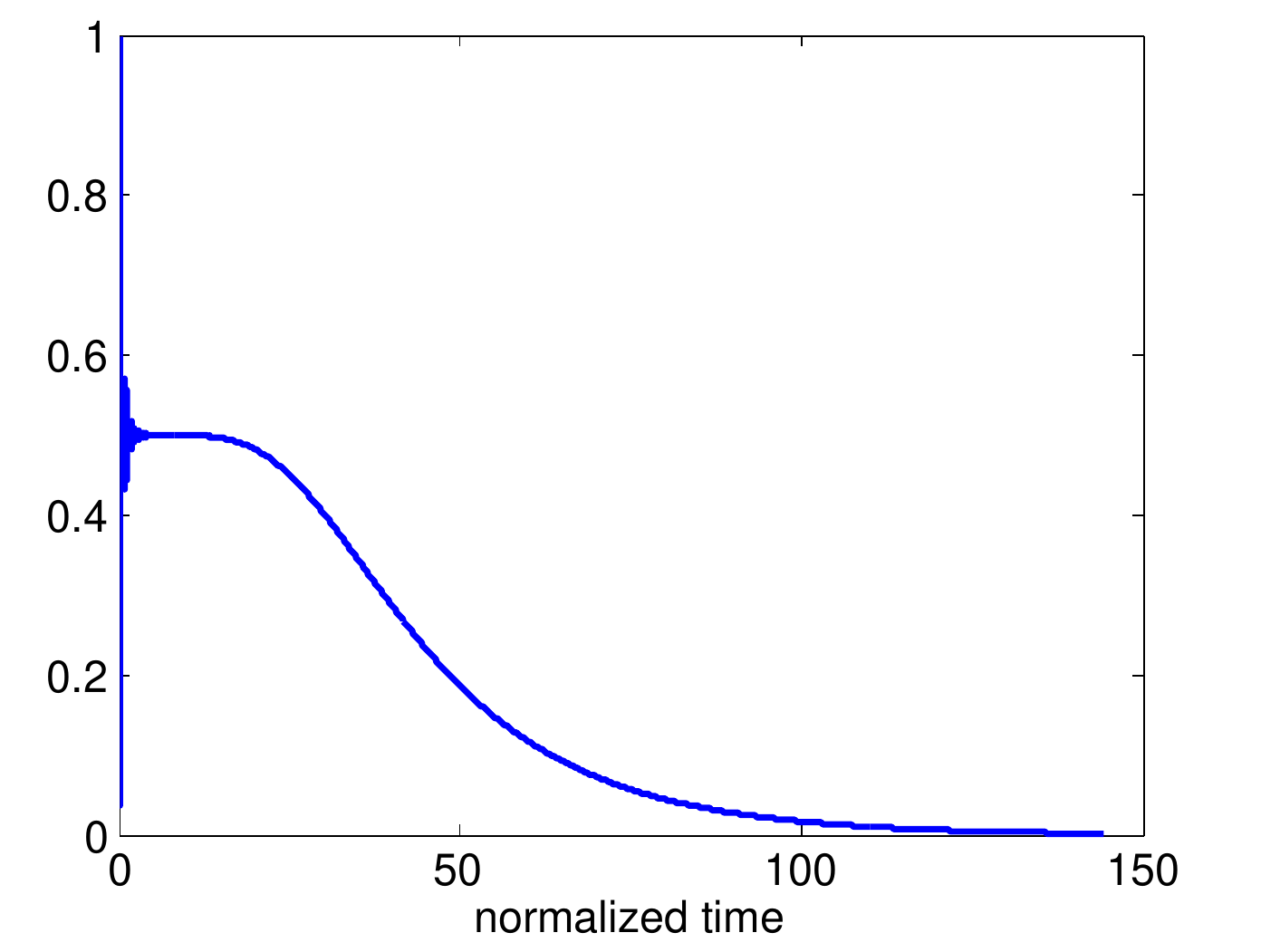}}
\centerline{\includegraphics[width=0.8\columnwidth]{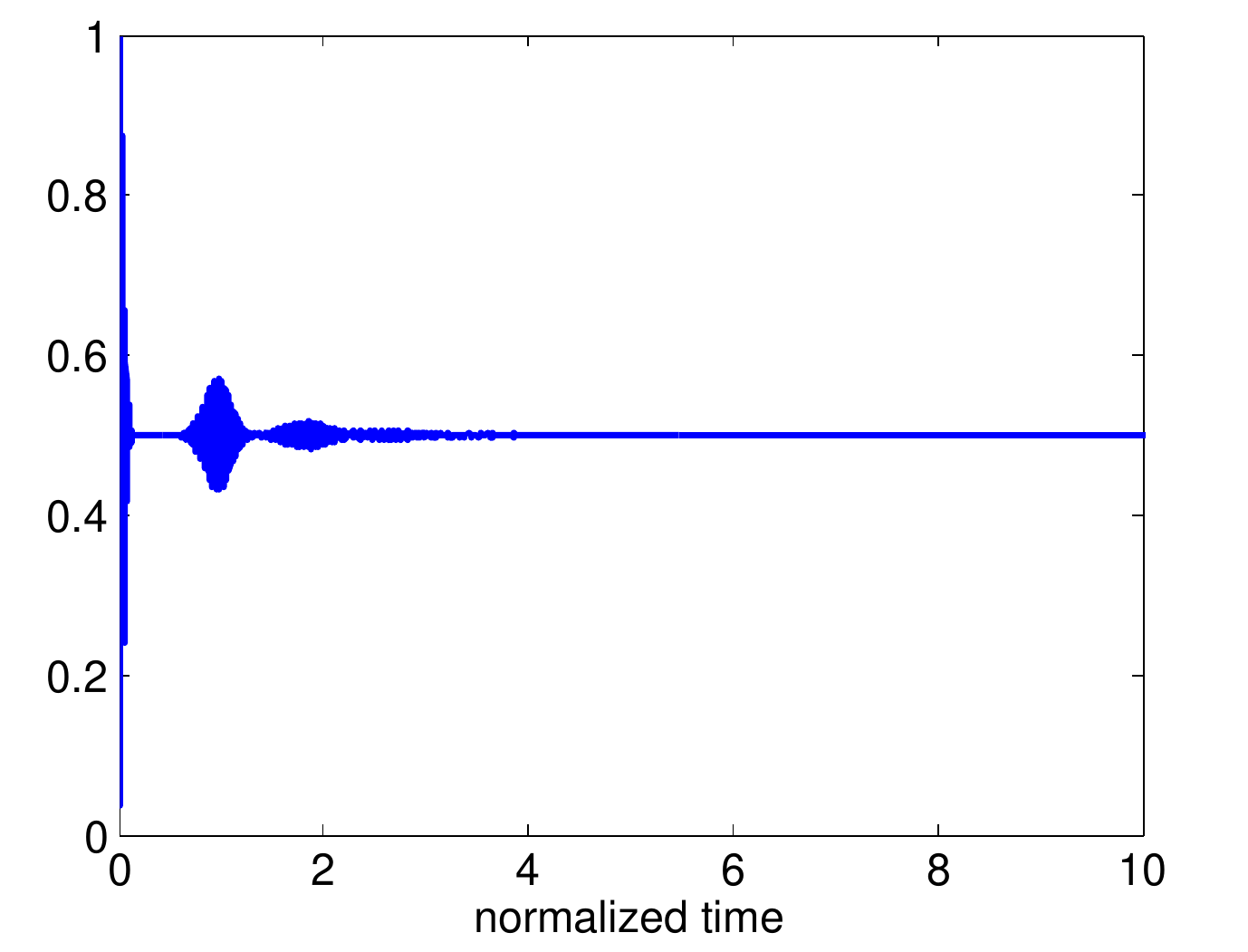}}
  \caption{(color online) High precision  full-rank numerical simulation of  $t\mapsto \rho(t)$  for the quantum collapse and revivals of a qubit resonantly coupled to a slightly damped quantum harmonic oscillator governed by~\eqref{eq:JCdamped}. Top plot: the solid curve corresponds to the excited population  $\bket{e|\rho|e}$ versus time; the normalized time corresponds to $t/T_r$ where $T_r$ is the  revival time; the oscillator damping time is around $10 T_r$; the initial   state  $\rho_0$ corresponds to the qubit in excited state $\ket{e}$ and oscillator in a coherent state with $\bar n=15$ photons;  a truncation of  the number of photons to a maximum of $2\bar n =30$ yields an  underlying  Hilbert space of dimension $62$.
  Bottom plot: zoom of  top plot for a normalized time  between $0$ and $10$.}  \label{fig:Exact}
\end{figure}

\begin{figure}
\centerline{\includegraphics[width=0.8\columnwidth]{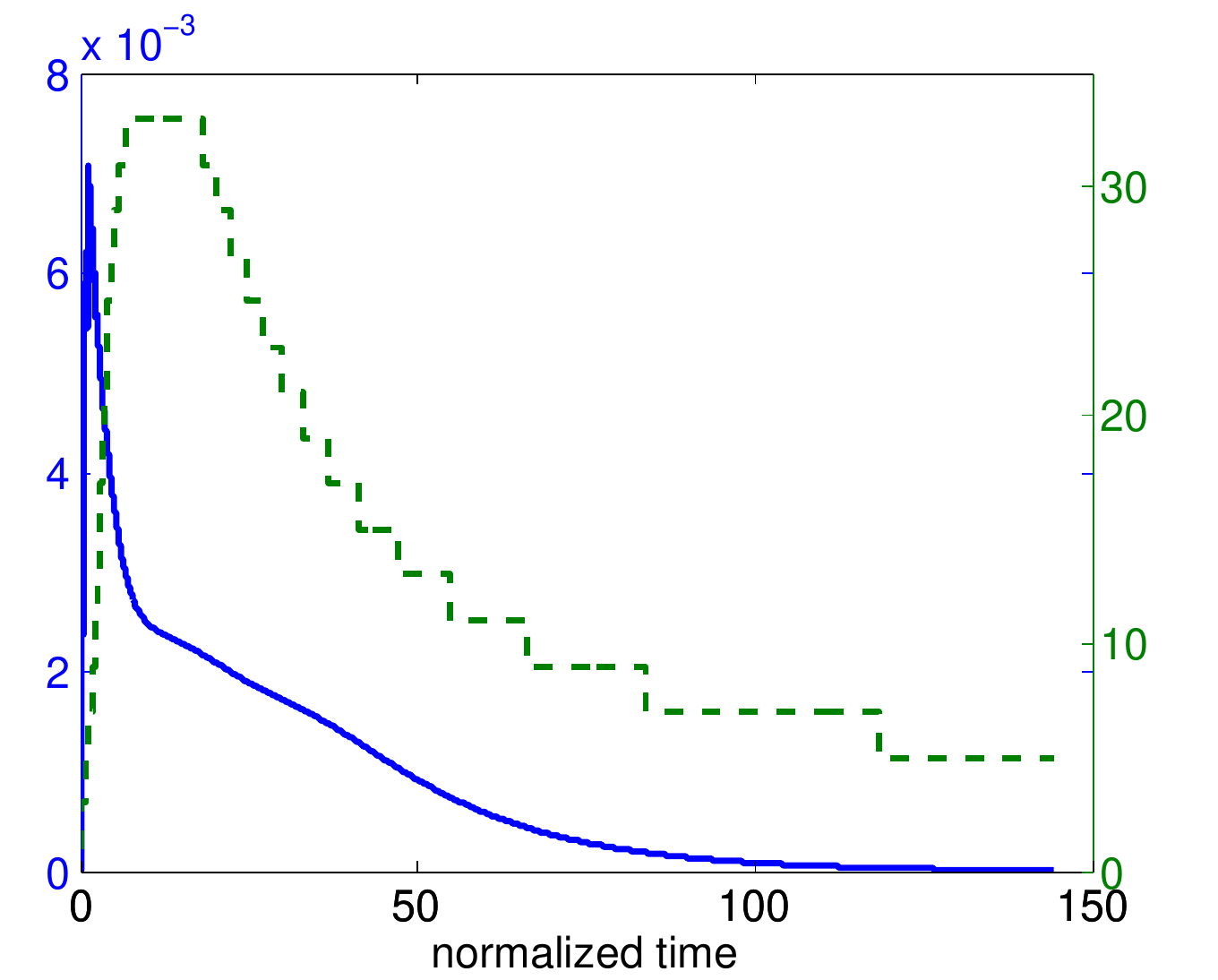}}
\centerline{\includegraphics[width=0.8\columnwidth]{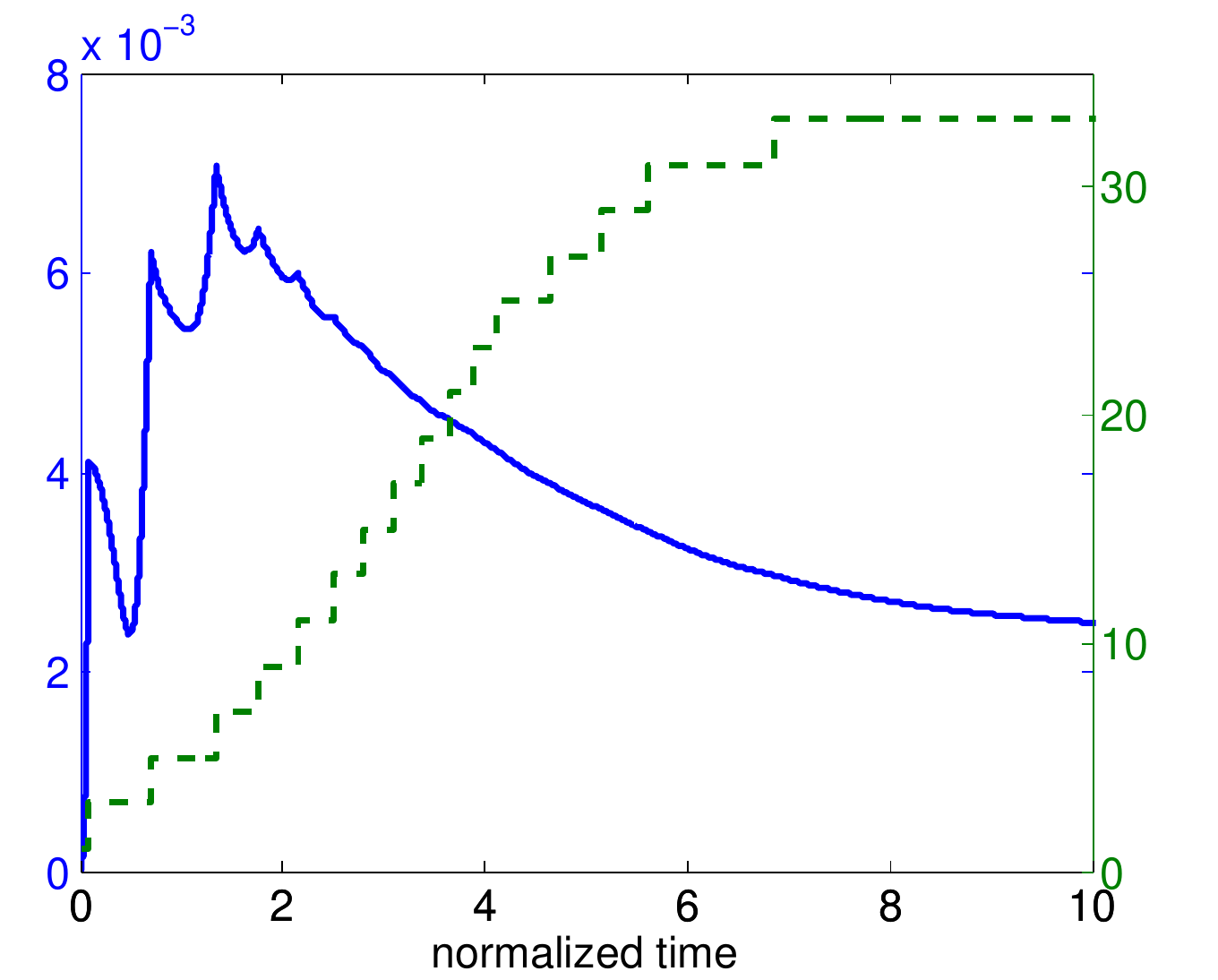}}
  \caption{(color online) Comparison between the full-rank  trajectory   $t\mapsto \rho(t)$ illustrated on   figure~\ref{fig:Exact} with the   adaptive rank trajectory  $t\mapsto\rhoLR(t)$ governed by~(\ref{eq:UH}-\ref{eq:sigH}). Top plot: the solid curve corresponds to $\sqrt{\tr{(\rho-\rhoLR)^2}}$ versus time  with scale on the left. The dashed curve corresponds to the adaptive rank $m$  with scale on the right;  the initial  low rank  state $\rhoLR$ coincides with $\rho_0$; the initial rank $m$ is  thus set to one; it evolves according to the maximal angular error $\theta_{max}=\frac{1}{1000}$.
  Bottom plot: zoom of  top plot for a normalized time  between $0$ and $10$.}  \label{fig:LRadapt}
\end{figure}

We have performed a comprehensive series of test of the approach. The practical implementation of the dynamical adaptivity of the rank is performed as follows. We denote by $\displaystyle \theta(\rhoLR) \equiv \frac{\| \mathcal{L}^\perp(\rhoLR) \|}{\| \mathcal{L}^\parallel(\rhoLR) \|}$ , fix a maximal angular error~$\theta_{max}$ and update the rank
\begin{itemize}
\item increasing $m$ by 1, when $\theta(\rhoLR) > \theta_{max}$   and   then complementing~$U$ via the solution of~\eqref{eq:V};
\item reducing $m$ by 1, when the smallest eigenvalue $\lambda_{min}$ of~$\sigma$ is such that  $\theta(\rhoLR) + \lambda_{min} < \frac 1 2 \theta_{max}$.
\end{itemize}
\medskip
As an illustrative example, we consider  a qubit  resonantly coupled to a quantized  harmonic damped  oscillator:
\begin{equation}\label{eq:JCdamped}
\dotex \rho = \frac{\Omega_0}{2} [ \ba^\dag \Sm - \ba \Sp, \rho]  - \kappa ( \bn \rho/2+\rho \bn/2 - \ba \rho \ba^\dag ),
\end{equation}
with  $\Omega_0 >0$ the vacuum Rabi pulsation, $\ba$ the photon  annihilation operator, $\Sm$ the qubit lowering operator ($\Sm=\GE$),  $\Sp=\Sm^\dag$,  $\bn=\ba^\dag \ba$ the  photon-number operator and $1/\kappa>0$ the  oscillator damping time.
In~Figure~3.20, page 156 of~\cite{haroche-raimondBook06},  numerical simulations of quantum  collapse and revivals are presented, for~$\kappa=0$, when the qubit is initially in the excited state $\ket{e}$ and the  oscillator in a coherent  state with $\bar n = 15$ photons. We consider here the same system with  $\kappa=\Omega_0/500$, a value small  compared to $\Omega_0$. This  corresponds to  a  photon life-time $1/\kappa$  around 10 times as large as the  revival time $T_{r}= \frac{4\pi\sqrt{\bar n}}{\Omega_0}$. In this  simple case, we can perform the full-rank simulation with high precision. This  simulation illustrated on Figure~\eqref{fig:Exact} provides us with a reference calculation.    As shown in~\cite{LeR2013PRA}, a constant  rank of 4 is sufficient to compute accurately the solution $t\mapsto \rho_t$ for $t$ between $0$ and $2 T_r$.  For intermediate values of~$t$, larger than~$2 T_r$ but not excessively larger,  the quantum state $\rho$  gets more and more mixed. For $t$ very large,  $\rho$ becomes again pure, since its limit for $t\mapsto+\infty$   is  the lowest  energy state  (qubit  in the ground state $\ket{g}$,  oscillator  with   zero photon). This behavior,  illustrated by the numerical simulations of Figure~\ref{fig:LRadapt},  is well captured by our adaptive approach.

\section{Denoised   Monte-Carlo approach }

In the case when the Lindblad equation~\eqref{eq:lindblad} is genuinely high-dimensional, the model reduction previously described is likely to be either ineffective or inaccurate, while the direct integration of the equation is out of reach. In this situation, the classical approach is to use a Monte-Carlo sampling. One derives a stochastic dynamics on the wave function $ \ket{\psi_t}$ such that the density $\rho(t)= \mathbb{E}\left( \ket{\psi_t}\bra{\psi_t}\right) $ constructed from $\ket{\psi_t}$ ($\mathbb{E}$ stands for expectation value, i.e. ensemble average) solves the Lindblad equation~\eqref{eq:lindblad}. In practice, $M$ independent trajectories,   $\ket{\psi_t^{(k)}}$ with $k=1,\ldots,M$,  are then simulated using that stochastic dynamics and the estimator of the mean
\begin{equation}
\label{eq:moyenne mc}
\bar{\rho}_{\rm{MC}}(t) = \frac 1 M \sum_{k=1}^M \ket{\psi_t^{(k)}}\bra{\psi_t^{(k)}}
\end{equation}
is used as an approximation of $\rho(t)$. The practical difficulty of Monte-Carlo approaches is, as briefly mentioned, above, the variance, that is, the noise intrinsically present in the approach.

 In principle, there are infinitely many dynamics on~$\ket{\psi_t}$ that are consistent with the Lindblad dynamics.
Interestingly, a straightforward calculation from~\eqref{eq:moyenne mc}  shows that
\begin{equation}\label{eq:MCvar}
\mathbb{E}\left( \tr{\left( \bar{\rho}_{\rm{MC}} - \rho \right)^2}\right) = \frac{1-\tr{\rho^2}}{M}
.
\end{equation}
The variance of the estimator of the mean $\bar{\rho}_{\rm{MC}}$ is therefore independent of  the specific unravelling choice, namely the stochastic dynamics set on~$ \ket{\psi_t}$, \emph{provided} that state $ \ket{\psi_t}$ remains normalized. This is easily seen in the proof of~\eqref{eq:MCvar}. This property is, of course, a remarkable peculiarity of the present context.  And the discretization in time of the process can of course slightly affect that property.  The classical unravelling choice (see~\cite{GisinP1992JoPAMaG,BreuerPetruccioneBook,BarchielliGregorattiBook}) is the Wiener process defined by
\begin{equation}
\label{eq:eds norm}
d  \ket{\psi_t} = D_1(\ket{\psi_t}) \,dt + D_2(\ket{\psi_t}) d {W}_t ,
\end{equation}
with the drift term
\begin{eqnarray}
\label{eq:def D10}
D_1(\ket{\psi}) &=& -i  H\ket{\psi} + D_1^0(\ket{\psi} )\quad\hbox{\rm where}\nonumber\\
D_1^0(\ket{\psi} )& =& \frac 1 2 \left( \langle L + L^{\dagger}\rangle_{\ket{\psi} } L - L^{\dagger} L - \frac 1 4 \langle L + L^{\dagger}\rangle^2_{\ket{\psi} } \right) \ket{\psi} ,\nonumber\\
\end{eqnarray}
and the diffusion
\begin{equation}
\label{eq:def D2}
D_2(\ket{\psi} ) = \left( L - \frac 1 2 \langle L + L^{\dagger}\rangle_{\ket{\psi} } \right)\ket{\psi} .
\end{equation}
We have used the notation $\langle A\rangle_{\ket{\psi} } = \langle \psi | A | \psi \rangle$.
We emphasize that other choices of dynamics on $\ket\psi$, all consistent with the Lindblad equation through~$\rho(t)= \mathbb{E}\left( \ket{\psi_t}\bra{\psi_t}\right) $, could be made. In particular, Poisson processes could be considered instead of Wiener processes (a choice that can be seen as more natural given the applications addressed in the present work where photons are emitted or absorbed). In any event, given the property~\eqref{eq:MCvar} and assuming that the Poisson process remains normalized, the variance remains identical. We indeed double-checked  that, in actuality, using Poisson processes does not bring further practical variance reduction, {\it see}~\cite{RousselMaster2015} for more details.

Variance is an issue for the numerical simulation and affects the accuracy of the results. It is thus desirable to come up with further, dedicated variance reduction approaches that may reduce the computational cost at accuracy fixed, or improve the accuracy for a given computational cost.
An approach that has proved effective in many engineering sciences for reducing variance of Monte-Carlo simulations is that of \emph{control variate}, see e.g.~\cite[page 54]{GrahamTalayBook2013}. In short, the approach consists in \emph{concurrently} simulating the original system under consideration and a system \emph{correlated} to that original system  so as to minimize the variance in the simulation of the former system. Intuitively, the approach works by "cancellation" of the noise because the same random draws are used for both systems. More specifically, we consider here as control variate  the   low-rank approximation $\rhoLR$  of  previous section. Even though that low-rank system  is not a correct approximation of the original system (which we have deliberately assumed here high-dimensional), it is sufficiently correlated to that system to provide an efficient variance reduction.
Practically, we construct an ("Control-Variate") estimated density depending on the adjustable scalar parameter $\lambda$,
\begin{equation}
	\label{eq:variable controle}
	\bar{\rho}_{\rm{CV}} = \bar \rho_{MC} + \lambda (\rho_{\rm{LR}} - \bar \rho_{MCLR}),
\end{equation}
as a combination of the original Monte-Carlo estimated  density~$\bar \rho_{MC}$ (constructed from the simulation of ~\eqref{eq:moyenne mc}-\eqref{eq:eds norm}), the estimated  density~$\bar \rho_{MCLR}$ constructed from a low-rank dynamics ("Monte-Carlo-Low-Rank", see below~\eqref{eq:eds mclr psi}), and   the density~$\rho_{\rm{LR}} $ obtained upon solving the corresponding low-rank Lindblad equation. Since by construction
$$\mathbb{E}[\bar{\rho}_{\rm{MCLR}}(t)] = \rho_{\rm{LR}}(t),$$
the approximation method~\eqref{eq:variable controle} is unbiased (taking the expectation of both sides of~\eqref{eq:variable controle}  yields $\rho=\mathbb{E}[\bar{\rho}_{\rm{CV}} ]=\mathbb{E}[ \bar \rho_{MC} ]$).
The scalar parameter $\lambda$ is adjusted so as to minimize the variance $\mathbb{E}\left(  \tr{\left( \bar{\rho}_{\rm{CV}} - \rho \right)^2}\right)$, a polynomial in $\lambda$ of degree 2, in such a way that
$$
\mathbb{E}\left(  \tr{\left( \bar{\rho}_{\rm{CV}} - \rho \right)^2}\right) \ll \mathbb{E}\left(  \tr{\left( \bar{\rho}_{\rm{MC}} - \rho \right)^2}\right)
$$
so that the simulation of $\bar{\rho}_{\rm{CV}}$ is eventually more effective than that of $\bar{\rho}_{\rm{MC}}$.  The more correlated the reduced model and the original model, the closer $\lambda$  to one, the smaller this  variance  and thus the more efficient the denoising.
In passing, we notice that, although this will not be the case in the actual numerical experiments we perform, the low-rank dynamics~$\rho_{\rm{LR}}(t)$ could itself be chosen with an adaptive rank, as in the previous section. Likewise, we could pick as control variate another dynamics than the low-rank dynamics, if a more convenient one is available.

The remaining question is to derive a stochastic dynamics on a wave function $\ket{\psiLR} $ such that  $\mathbb{E}\left(\ket{\psiLR}\bra{\psiLR}\right)=\rhoLR$,
\emph{given} that the dynamics of~$\rhoLR$ is known since easy to compute via $(U,\sigma) $ solutions of~\eqref{eq:UH} and~\eqref{eq:sigH}. For this purpose, it is a natural idea to seek~$\ket{\psiLR}$ under the form~$\ket{\psiLR}= U \ket{\nu} $ where the  reduced  wave function $\ket{\nu} \in \mathbb{R}^m$ is a stochastic process to be determined. It can be shown (see~\cite{RousselMaster2015}) that the correct dynamics to consider reads
\begin{multline}
\label{eq:eds mclr psi}
	d  \ket{\psiLR} = P_\rhoLR D_1(\ket{\psiLR}) \,dt + P_\rhoLR D_2(\ket{\psiLR}) d  W_t \\
	+ (\mathbb{I}_n-P_\rhoLR) \left(-\frac 1 2 L^{\dagger} L + LU \sigma U^{\dagger} L^{\dagger} U \sigma^{-1} U^{\dagger} \right)  \ket{\psiLR}  \,dt \\
	+ \frac {\tr{(\mathbb{I}_n-P_\rhoLR)L \rhoLR L^{\dagger}}}{2m} U \sigma^{-1} U^{\dagger} \ket{\psiLR} \,dt, \\
\end{multline}
with  $D_1$  and $D_2$ defined in~\eqref{eq:def D10}  and~\eqref{eq:def D2}.

\begin{figure}
\centerline{\includegraphics[width=0.8\columnwidth]{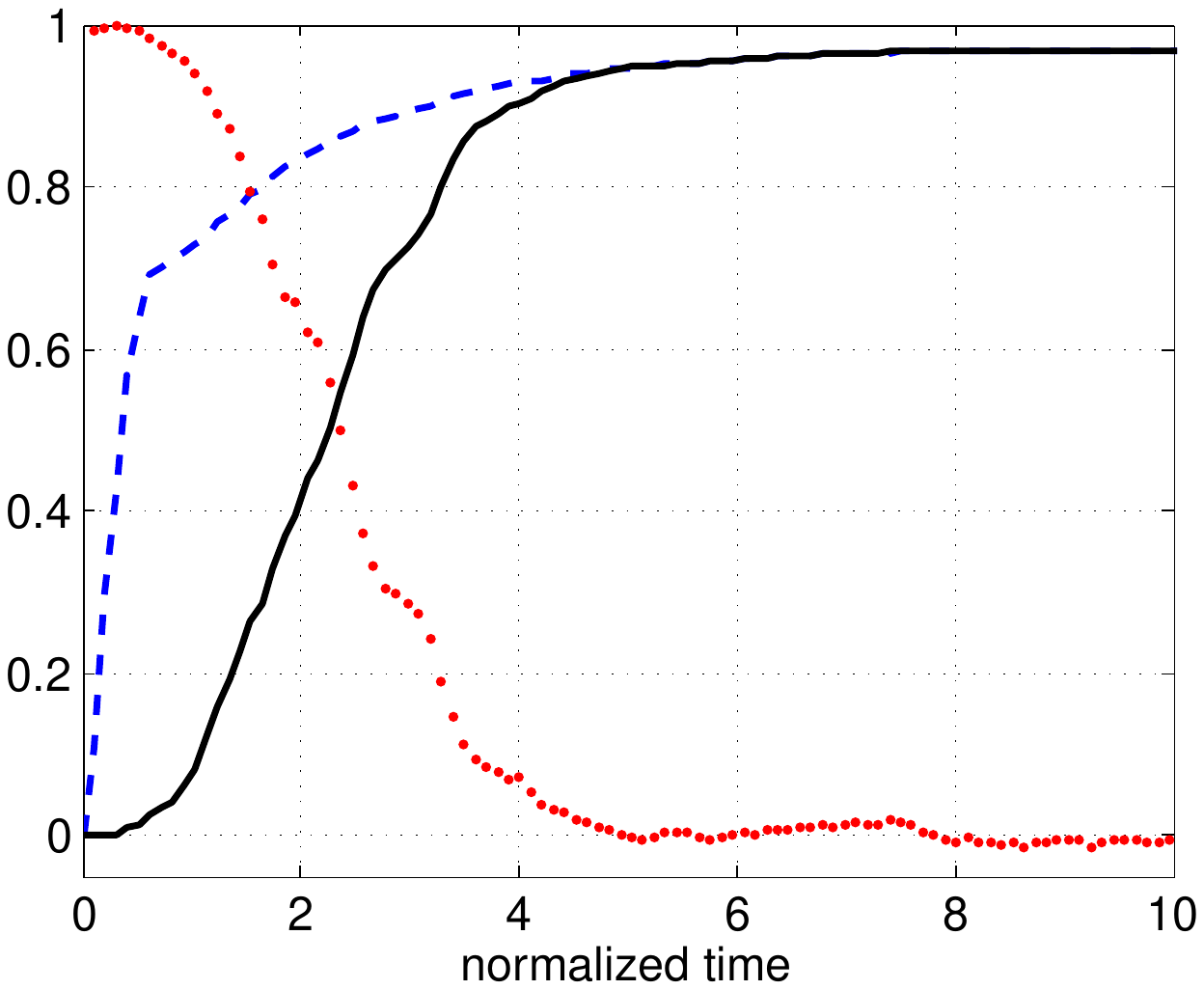}}
  \caption{(color online) Variance reduction corresponding to the  quantum collapse and revivals  in  the bottom plot of Figure~\ref{fig:LRadapt}. The solid black curve corresponds to  $\sqrt{\tr{(\rho-\bar{\rho}_{\rm{CV}})^2}}$, the  dashed blue  one  to $\sqrt{\tr{(\rho-\bar{\rho}_{\rm{MC}})^2}}$ and the dotted red one to $\lambda$. The number of trajectories $M$ is  set to $400$. The low-rank approximation~$\rhoLR$ used for the control variate governed by~\eqref{eq:eds mclr psi} is maintained at the  constant rank $m=2$. }\label{fig:CV10}
\end{figure}

 Since $\bar{\rho}_{\rm{MCLR}} = \frac 1 M \sum_{k=1}^M \ket{\psiLR^{(k)}}\bra{\psiLR^{(k)}}$, a simple computation,  exploiting the fact that  both  $\ket{\psi^{(k)}}$ and $\ket{\psiLR^{(k)}}$ are normalized ,  shows that
\begin{multline*}
  M ~ \mathbb{E}\left(  \tr{\left( \bar{\rho}_{\rm{CV}} - \rho \right)^2}\right) = \left(1-\tr{\rho_{\text{\tiny LR}}^2}\right) \lambda^2
  \\
  + 2  \left(\mathbb{E}\left(\left| \langle \psi | \psiLR \rangle\right| ^2\right)- \tr{\rho\rhoLR} \right) \lambda
  +1-\tr{\rho^2}
  .
\end{multline*}
Thus the optimal  denoising choice  for  $\lambda$ reads
\begin{equation} \label{eq:lambda}
  \lambda= \frac{\tr{\rho\rhoLR} - \mathbb{E}\left(\left| \langle \psi | \psiLR \rangle \right| ^2\right)}{1-\tr{\rhoLR^2}}
  .
\end{equation}
In practice,   {\it and also} in the simulations of  Figure~\eqref{fig:CV10}, the adjustable parameter $\lambda$ is given by~\eqref{eq:lambda} where the  $\rho$ is replaced by  $\bar{\rho}_{\rm{MC}}$ (we do not have access to $\rho$ itself) and where $\mathbb{E}\left(\left| \langle \psi | \psiLR \rangle \right| ^2\right)$ is replaced by the  estimated  mean $\frac 1 M \sum_{k=1}^M \left| \langle \psi ^{(k)}| \psiLR^{(k)} \rangle \right| ^2$.

Figure~\eqref{fig:CV10} illustrates the interest of this denoising method for the Lindblad system  simulated  in Figure~\eqref{fig:LRadapt}. We take $M=400$  trajectories. The  dynamics of   $\ket{\psiLR}$ is based on a low-rank approximation  $\rhoLR$ of constant rank $m=2$. On the bottom plot of  Figure~\ref{fig:LRadapt}, we observe that, at normalized  time  2, an accurate approximation of $\rho$  must be of rank  $10$ or  larger. Nevertheless,  Figure~\ref{fig:CV10} indicates that, at  normalized time 2,  a reduction of the standard deviation   around  50\% is obtained with $\bar{\rho}_{\rm{CV}}$ instead of $\bar{\rho}_{\rm{MC}}$.  Such a variance reduction corresponds,  with a classical quantum Monte-Carlo method,  to an  increase  of the number  of trajectories  by a factor  $4$.  Such a gain confirms the definite interest of  combining deterministic low-rank approximations with quantum  Monte-Carlo trajectories for the numerical simulation of high-dimensional Lindblad equation.

\section*{Acknowledgments}


The first two authors are partially supported by the ANR, Projet Blanc EMAQS ANR-2011-BS01-017-01.


\begin{thebibliography}{14}%
\makeatletter
\providecommand \@ifxundefined [1]{%
 \@ifx{#1\undefined}
}%
\providecommand \@ifnum [1]{%
 \ifnum #1\expandafter \@firstoftwo
 \else \expandafter \@secondoftwo
 \fi
}%
\providecommand \@ifx [1]{%
 \ifx #1\expandafter \@firstoftwo
 \else \expandafter \@secondoftwo
 \fi
}%
\providecommand \natexlab [1]{#1}%
\providecommand \enquote  [1]{``#1''}%
\providecommand \bibnamefont  [1]{#1}%
\providecommand \bibfnamefont [1]{#1}%
\providecommand \citenamefont [1]{#1}%
\providecommand \href@noop [0]{\@secondoftwo}%
\providecommand \href [0]{\begingroup \@sanitize@url \@href}%
\providecommand \@href[1]{\@@startlink{#1}\@@href}%
\providecommand \@@href[1]{\endgroup#1\@@endlink}%
\providecommand \@sanitize@url [0]{\catcode `\\12\catcode `\$12\catcode
  `\&12\catcode `\#12\catcode `\^12\catcode `\_12\catcode `\%12\relax}%
\providecommand \@@startlink[1]{}%
\providecommand \@@endlink[0]{}%
\providecommand \url  [0]{\begingroup\@sanitize@url \@url }%
\providecommand \@url [1]{\endgroup\@href {#1}{\urlprefix }}%
\providecommand \urlprefix  [0]{URL }%
\providecommand \Eprint [0]{\href }%
\providecommand \doibase [0]{http://dx.doi.org/}%
\providecommand \selectlanguage [0]{\@gobble}%
\providecommand \bibinfo  [0]{\@secondoftwo}%
\providecommand \bibfield  [0]{\@secondoftwo}%
\providecommand \translation [1]{[#1]}%
\providecommand \BibitemOpen [0]{}%
\providecommand \bibitemStop [0]{}%
\providecommand \bibitemNoStop [0]{.\EOS\space}%
\providecommand \EOS [0]{\spacefactor3000\relax}%
\providecommand \BibitemShut  [1]{\csname bibitem#1\endcsname}%
\let\auto@bib@innerbib\@empty
\bibitem [{\citenamefont {Dalibard}\ \emph {et~al.}(1992)\citenamefont
  {Dalibard}, \citenamefont {Castin},\ and\ \citenamefont
  {M{\o}lmer}}]{DalibCM1992PRL}%
  \BibitemOpen
  \bibfield  {author} {\bibinfo {author} {\bibfnamefont {J.}~\bibnamefont
  {Dalibard}}, \bibinfo {author} {\bibfnamefont {Y.}~\bibnamefont {Castin}}, \
  and\ \bibinfo {author} {\bibfnamefont {K.}~\bibnamefont {M{\o}lmer}},\ }\href
  {http://link.aps.org/doi/10.1103/PhysRevLett.68.580} {\bibfield  {journal}
  {\bibinfo  {journal} {Phys. Rev. Lett.}\ }\textbf {\bibinfo {volume} {68}},\
  \bibinfo {pages} {580} (\bibinfo {year} {1992})}\BibitemShut {NoStop}%
\bibitem [{\citenamefont {Le~Bris}\ and\ \citenamefont
  {Rouchon}(2013)}]{LeR2013PRA}%
  \BibitemOpen
  \bibfield  {author} {\bibinfo {author} {\bibfnamefont {C.}~\bibnamefont
  {Le~Bris}}\ and\ \bibinfo {author} {\bibfnamefont {P.}~\bibnamefont
  {Rouchon}},\ }\href {http://link.aps.org/doi/10.1103/PhysRevA.87.022125}
  {\bibfield  {journal} {\bibinfo  {journal} {Phys. Rev. A}\ }\textbf {\bibinfo
  {volume} {87}},\ \bibinfo {pages} {022125} (\bibinfo {year}
  {2013})}\BibitemShut {NoStop}%
\bibitem [{\citenamefont {Handel}\ and\ \citenamefont
  {Mabuchi}(2005)}]{Handel-mabuchi:JOB2005}%
  \BibitemOpen
  \bibfield  {author} {\bibinfo {author} {\bibfnamefont {R.~v.}\ \bibnamefont
  {Handel}}\ and\ \bibinfo {author} {\bibfnamefont {H.}~\bibnamefont
  {Mabuchi}},\ }\href {http://stacks.iop.org/1464-4266/7/i=10/a=005} {\bibfield
   {journal} {\bibinfo  {journal} {Journal of Optics B: Quantum and
  Semiclassical Optics}\ }\textbf {\bibinfo {volume} {7}},\ \bibinfo {pages}
  {S226} (\bibinfo {year} {2005})}\BibitemShut {NoStop}%
\bibitem [{\citenamefont {Mabuchi}(2008)}]{Mabuchi:PRA2008}%
  \BibitemOpen
  \bibfield  {author} {\bibinfo {author} {\bibfnamefont {H.}~\bibnamefont
  {Mabuchi}},\ }\href {\doibase 10.1103/PhysRevA.78.015801} {\bibfield
  {journal} {\bibinfo  {journal} {Phys. Rev. A}\ }\textbf {\bibinfo {volume}
  {78}},\ \bibinfo {pages} {015801} (\bibinfo {year} {2008})}\BibitemShut
  {NoStop}%
\bibitem [{\citenamefont {Lubich}(2008)}]{MR2474331}%
  \BibitemOpen
  \bibfield  {author} {\bibinfo {author} {\bibfnamefont {C.}~\bibnamefont
  {Lubich}},\ }\href {\doibase 10.4171/067} {\emph {\bibinfo {title} {From
  quantum to classical molecular dynamics: reduced models and numerical
  analysis}}},\ Zurich Lectures in Advanced Mathematics\ (\bibinfo  {publisher}
  {European Mathematical Society (EMS), Z\"urich},\ \bibinfo {year} {2008})\
  pp.\ \bibinfo {pages} {x+144}\BibitemShut {NoStop}%
\bibitem [{\citenamefont {Lubich}\ and\ \citenamefont
  {Oseledets}(2014)}]{LubichBIT2014}%
  \BibitemOpen
  \bibfield  {author} {\bibinfo {author} {\bibfnamefont {C.}~\bibnamefont
  {Lubich}}\ and\ \bibinfo {author} {\bibfnamefont {I.~V.}\ \bibnamefont
  {Oseledets}},\ }\href {\doibase 10.1007/s10543-013-0454-0} {\bibfield
  {journal} {\bibinfo  {journal} {BIT}\ }\textbf {\bibinfo {volume} {54}},\
  \bibinfo {pages} {171} (\bibinfo {year} {2014})}\BibitemShut {NoStop}%
\bibitem [{\citenamefont {M{\o}lmer}\ \emph {et~al.}(1993)\citenamefont
  {M{\o}lmer}, \citenamefont {Castin},\ and\ \citenamefont
  {Dalibard}}]{Moelmer1993}%
  \BibitemOpen
  \bibfield  {author} {\bibinfo {author} {\bibfnamefont {K.}~\bibnamefont
  {M{\o}lmer}}, \bibinfo {author} {\bibfnamefont {Y.}~\bibnamefont {Castin}}, \
  and\ \bibinfo {author} {\bibfnamefont {J.}~\bibnamefont {Dalibard}},\ }\href
  {http://josab.osa.org/abstract.cfm?URI=josab-10-3-524} {\bibfield  {journal}
  {\bibinfo  {journal} {J. Opt. Soc. Am. B}\ }\textbf {\bibinfo {volume}
  {10}},\ \bibinfo {pages} {524} (\bibinfo {year} {1993})}\BibitemShut
  {NoStop}%
\bibitem [{\citenamefont {Castin}\ and\ \citenamefont
  {M{\o}lmer}(1995)}]{CastiM1995PRL}%
  \BibitemOpen
  \bibfield  {author} {\bibinfo {author} {\bibfnamefont {Y.}~\bibnamefont
  {Castin}}\ and\ \bibinfo {author} {\bibfnamefont {K.}~\bibnamefont
  {M{\o}lmer}},\ }\href {http://link.aps.org/doi/10.1103/PhysRevLett.74.3772}
  {\bibfield  {journal} {\bibinfo  {journal} {Phys. Rev. Lett.}\ }\textbf
  {\bibinfo {volume} {74}},\ \bibinfo {pages} {3772} (\bibinfo {year}
  {1995})}\BibitemShut {NoStop}%
\bibitem [{\citenamefont {Haroche}\ and\ \citenamefont
  {Raimond}(2006)}]{haroche-raimondBook06}%
  \BibitemOpen
  \bibfield  {author} {\bibinfo {author} {\bibfnamefont {S.}~\bibnamefont
  {Haroche}}\ and\ \bibinfo {author} {\bibfnamefont {J.}~\bibnamefont
  {Raimond}},\ }\href@noop {} {\emph {\bibinfo {title} {Exploring the Quantum:
  Atoms, Cavities and Photons.}}}\ (\bibinfo  {publisher} {Oxford University
  Press},\ \bibinfo {year} {2006})\BibitemShut {NoStop}%
\bibitem [{\citenamefont {Roussel}(2015)}]{RousselMaster2015}%
  \BibitemOpen
  \bibfield  {author} {\bibinfo {author} {\bibfnamefont {J.}~\bibnamefont
  {Roussel}},\ }\emph {\bibinfo {title} {{Numerical simulation of high
  dimensional open quantum systems (in French)}}},\ \href
  {https://hal.inria.fr/hal-01205747} {Master's thesis},\ \bibinfo  {school}
  {{Univ. Pierre et Marie Curie, Laboratoire J.L. Lions},
  https://hal.inria.fr/hal-01205747} (\bibinfo {year} {2015})\BibitemShut
  {NoStop}%
\bibitem [{\citenamefont {Gisin}\ and\ \citenamefont
  {Percival}(1992)}]{GisinP1992JoPAMaG}%
  \BibitemOpen
  \bibfield  {author} {\bibinfo {author} {\bibfnamefont {N.}~\bibnamefont
  {Gisin}}\ and\ \bibinfo {author} {\bibfnamefont {I.~C.}\ \bibnamefont
  {Percival}},\ }\href {http://stacks.iop.org/0305-4470/25/i=21/a=023}
  {\bibfield  {journal} {\bibinfo  {journal} {Journal of Physics A:
  Mathematical and General}\ }\textbf {\bibinfo {volume} {25}},\ \bibinfo
  {pages} {5677} (\bibinfo {year} {1992})}\BibitemShut {NoStop}%
\bibitem [{\citenamefont {Breuer}\ and\ \citenamefont
  {Petruccione}(2006)}]{BreuerPetruccioneBook}%
  \BibitemOpen
  \bibfield  {author} {\bibinfo {author} {\bibfnamefont {H.-P.}\ \bibnamefont
  {Breuer}}\ and\ \bibinfo {author} {\bibfnamefont {F.}~\bibnamefont
  {Petruccione}},\ }\href@noop {} {\emph {\bibinfo {title} {The Theory of Open
  Quantum Systems}}}\ (\bibinfo  {publisher} {{Clarendon-Press, Oxford}},\
  \bibinfo {year} {2006})\BibitemShut {NoStop}%
\bibitem [{\citenamefont {Barchielli}\ and\ \citenamefont
  {Gregoratti}(2009)}]{BarchielliGregorattiBook}%
  \BibitemOpen
  \bibfield  {author} {\bibinfo {author} {\bibfnamefont {A.}~\bibnamefont
  {Barchielli}}\ and\ \bibinfo {author} {\bibfnamefont {M.}~\bibnamefont
  {Gregoratti}},\ }\href@noop {} {\emph {\bibinfo {title} {Quantum Trajectories
  and Measurements in Continuous Time: the Diffusive Case}}}\ (\bibinfo
  {publisher} {Springer Verlag},\ \bibinfo {year} {2009})\BibitemShut {NoStop}%
\bibitem [{\citenamefont {Graham}\ and\ \citenamefont
  {Talay}(2013)}]{GrahamTalayBook2013}%
  \BibitemOpen
  \bibfield  {author} {\bibinfo {author} {\bibfnamefont {C.}~\bibnamefont
  {Graham}}\ and\ \bibinfo {author} {\bibfnamefont {D.}~\bibnamefont {Talay}},\
  }\href {\doibase 10.1007/978-3-642-39363-1} {\emph {\bibinfo {title}
  {Stochastic simulation and {M}onte {C}arlo methods}}},\ \bibinfo {series}
  {Stochastic Modelling and Applied Probability}, Vol.~\bibinfo {volume} {68}\
  (\bibinfo  {publisher} {Springer, Heidelberg},\ \bibinfo {year} {2013})\ pp.\
  \bibinfo {pages} {xvi+260}\BibitemShut {NoStop}%
\end{thebibliography}
%

\end{document}